\documentclass[fleqn,usenatbib]{mnras}
\usepackage{amsmath}    
\usepackage{newtxtext,newtxmath}
\usepackage[T1]{fontenc}
\usepackage{ae,aecompl}
\usepackage{graphicx}	
\usepackage{geometry}
\usepackage{float}
\usepackage{caption}
\usepackage{subcaption}
\usepackage{dblfloatfix}
\usepackage{url}
\usepackage{xcolor,colortbl}

\defcitealias{yang18}{yang2018}

\newcommand{\hmpc}{\,h^{-1}\,{\rm Mpc}}

\newcommand{\hmsun}{\,h^{-1}\,{\rm M}_\odot}

\title{Connection Between SDSS Galaxies and ELUCID Subhaloes in the Eye of Machine Learning}

\author[X. Xu et al.]{
Xiaoju Xu,$^{1}$\thanks{E-mail: xiaoju@sjtu.edu.cn}
Xiaohu Yang,$^{1,2}$
Haojie Xu$^{1,3}$
and Youcai Zhang$^{3}$\\
$^{1}$Department of Astronomy, School of Physics and Astronomy, and Shanghai Key Laboratory for Particle Physics and Cosmology, Shanghai Jiao Tong University, \\Shanghai 200240, People's Republic of China\\
$^{2}$Tsung-Dao Lee Institute, and Key Laboratory for Particle Physics, Astrophysics and Cosmology, Ministry of Education, Shanghai Jiao Tong University, \\Shanghai 201210, People's Republic of China\\
$^{3}$Shanghai Astronomical Observatory (SHAO), Nandan Road 80, Shanghai 200240, China}

\date{Accepted XXX. Received YYY; in original form ZZZ}

\pubyear{2023}

\begin{document}
\label{firstpage}
\pagerange{\pageref{firstpage}--\pageref{lastpage}}
\maketitle

\begin{abstract}
We explore the feasibility of learning the connection between SDSS galaxies and ELUCID subhaloes with random forest (RF). ELUCID is a constrained $N$-body simulation constructed using the matter density field of SDSS. Based on an SDSS-ELUCID matched catalogue, we build RF models that predict $M_r$ magnitude, colour, stellar mass $M_*$, and specific star formation rate (sSFR) with several subhalo properties. While the RF can predict $M_r$ and $M_*$ with reasonable accuracy, the prediction accuracy of colour and sSFR is low, which could be due to the mismatch between galaxies and subhaloes. To test this, we shuffle the galaxies in subhaloes of narrow mass bins in the local neighbourhood using galaxies of a semi-analytic model (SAM) and the TNG hydrodynamic simulation. We find that the shuffling only slightly reduces the colour prediction accuracy in SAM and TNG, which is still considerably higher than that of the SDSS. This suggests that the true connection between SDSS colour and subhalo properties could be weaker than that in the SAM and TNG without the mismatch effect. We also measure the Pearson correlation coefficient between galaxy properties and the subhalo properties in SDSS, SAM, and TNG. Similar to the RF results, we find that the colour-subhalo correlation in SDSS is lower than both the SAM and TNG. We also show that the galaxy-subhalo correlations depend on subhalo mass in the galaxy formation models. Advanced surveys with more fainter galaxies will provide new insights into the galaxy-subhalo relation in the real Universe.

\end{abstract}


\section{Introduction} 
\label{sec:intro}
Understanding the formation and evolution of galaxies is a crucial aspect of modern cosmology. In recent years, large-volume galaxy surveys such as Sloan Digital Sky Survey (SDSS, \citealt{York2000}), SDSS-III \citep{Eisenstein2011} and SDSS-IV \citep{Dawson2016}, and the Dark Energy Spectroscopic Instrument (DESI, \citealt{desi2016}) provide high-precision measurements of galaxy observables, leading to significant progress in this field. Since galaxies are believed to form within dark matter haloes, studying the connection between them can provide valuable insights into galaxy formation and evolution. However, unlike galaxy properties such as magnitude and colour which can be observed directly, the inner structure and formation histories of dark matter haloes are challenging to measure through observations.

In contrast, the formation history of dark matter halo and subhalo can be easily traced through $N$-body simulations, which evolve dark matter particles under gravity \citep{Springel2005, Prada2012, Wang2020}. To simulate galaxies, semi-analytic models (SAM) of galaxy formation processes can be implemented on the subhalo merger tree extracted from $N$-body simulations \citep{Guo2011, Guo2013, Croton2016, Cora2018}. Furthermore, hydrodynamic simulations are developed to produce galaxies in dark matter haloes by adding baryonic particles beyond dark matter particles \citep{Vogelsberger2014, Schaye2015, Nelson2015, Nelson2019}. Both SAM and hydrodynamic simulations can be tuned to reproduce statistical galaxy observables such as abundance and clustering. However, since the galaxy formation processes are not yet fully understood, simulated galaxies may still deviate from those in the real Universe. Additionally, it is difficult to compare simulated galaxies individually with the real ones, as the one-to-one correspondence between them is not guaranteed. 

One approach to address these issues is to construct constrained simulations based on the observed distribution of galaxies in the local universe. Using the group catalogue built from SDSS galaxies \citep{Yang2007,Yang2012}, the matter density field at low redshift can be constructed and treated as the final output of the constrained simulations \citep{Wang2009,Wang2012}. To infer the initial condition of the final density field, \citet{Wang2014} proposed a method that utilises the Hamiltonian Markov Chain Monte Carlo algorithm to sample the posterior distribution of the initial condition, together with a Particle-Mesh model that evolves the initial condition to the final state. With the constrained initial condition, \citet{Wang2016} carry out the ELUCID $N$-body simulation, which accurately reproduces the observed large-scale structures in SDSS Data Release 7 (DR7, \citealt{Abazajian2009}). Based on this similarity, \citet{Yang2018} implements a neighbourhood abundance matching method that matches the observed galaxies in DR7 to the subhaloes in the ELUCID simulation.

The one-to-one matching between observed galaxies and simulated subhaloes provides a novel path for investigating the galaxy-halo relation. It is shown that this approach can recover the massive haloes to a large extent \citep{Tweed2017}, and the haloes linked to the bright galaxies may represent the actual haloes in the Universe with a high possibility. This provides an opportunity to compare galaxies in observation with those in the SAM implemented on the ELUCID simulation and the upcoming ELUCID hydrodynamic simulation on an individual level. Such comparison is helpful for understanding the differences between galaxy formation models and the actual galaxy formation processes in the Universe. In addition, studying the galaxy-halo relation of the SDSS-ELUCID matching pairs statistically also provides insights into galaxy formation and evolution in the real Universe. In this work, we aim to capture this relation with machine learning and predicting galaxy properties based on subhalo properties.

Machine learning models are widely used in cosmological studies in the literature due to the ability to efficiently learn non-linear multi-variate dependencies between input and output variables. Efforts have been made on predicting halo occupations or galaxy properties with dark matter halo or subhalo properties based on SAM or hydrodynamic simulations \citep{Kamdar2016a, Kamdar2016b, Agarwal2018, Lovell2022, Xu2021b, Xu2022}.  Once trained, these machine learning models can be applied to large-volume $N$-body simulations to create mock galaxy catalogues that reproduce the galaxy-halo connection in corresponding galaxy formation models. In this work, we focus on predicting galaxy properties from subhalo properties based on the SDSS-ELUCID matching catalogue in \citet{Yang2018}, and we evaluate the feasibility of using machine learning to produce realistic mock catalogues with large-volume $N$-body simulations. However, the accuracy of the dark matter halo reconstruction in ELUCID, particularly for low-mass haloes, is not guaranteed, which may affect the robustness of our analysis. Therefore, we perform tests to estimate the impact of uncertainties in subhalo properties (or in other words, the mismatching between subhaloes and galaxies) on the prediction of galaxy properties. This study is helpful for revealing discrepancies between observed galaxies and modeled galaxies and shedding light on galaxy-subhalo relation in the real Universe. 

The structure of this paper is as follows. We provide an overview of the ELUCID $N$-body simulation, the SDSS-ELUCID matching catalogue, galaxy formation models, and the machine learning method we implemented in Section~\ref{sec:data}. The main results of predicting SDSS galaxy properties are shown in Section~\ref{sec:sdsspred}. We then investigate the possible effect of mismatching between galaxies and subhaloes with a SAM implemented on the ELUCID simulation and a hydrodynamic simulation in Section~\ref{sec:mismatch}. Finally, we summarise and discuss our results in Section~\ref{sec:summary}. 

\section{Data and methods} 
\label{sec:data}

\subsection{ELUCID simulation and SDSS-ELUCID matching catalogue} \label{sec:simu}

In this study, we utilise the SDSS-ELUCID matching catalogue from \citet{Yang2018} (Match2 method), which links observed galaxies to subhaloes in the ELUCID $N$-body simulation. ELUCID is a constrained simulation designed to reproduce the large-scale distributions of galaxies observed in the Northern Galactic Cap (NGC) region of SDSS DR7 \citep{Abazajian2009}, in the range of $99^\circ <$ R.A. $< 283^\circ$, $-7^\circ < $ dec. $<75^\circ$ and $0.01<z<0.12$. To achieve this, the matter density field reconstructed from the \citet{Yang2007} group catalogue, which is built based on the New York University Value-Added Galaxy Catalogue (NYU-VAGC, \citealt{Blanton2005}), is used as the final condition for inferring the corresponding initial condition. For this purpose, Hamiltonian Markov Chain Monte Carlo method (HMCMC, \citealt{Duane1987}) and PM dynamics \citep{White1983, Jing2002} are used. The former samples the posterior distribution of linear initial conditions with a specific final condition, and the latter evolves initial condition to final density field by efficiently evaluating gravitational forces at each time step. With the inferred initial condition, the ELUCID simulation evolves $3072^3$ dark matter particles of mass $3.0875 \times 10^8 \hmsun$ in a box with a comoving length of 500 $h^{-1}{\rm Mpc}$ on a side using an updated version of the GADGET-2 code \citep{Springel2005}. The simulation adopts the {\it WMAP5} cosmology with cosmological parameters $\Omega_{\rm m}=0.258$, $\Omega_{\rm b}=0.044$, $h=0.72$, and $n_s$=0.963, and $\sigma_8=0.796$ \citep{Dunkley2009}. 

In each snapshot of the simulation, dark matter haloes and subhaloes are identified using the Friend-of-Friend (FOF) algorithm \citep {Davis1985})  and \texttt{SUBFIND} method \citep{Springel2001}, respectively. Subhalo merger tree is then constructed by linking subhaloes from \texttt{SUBFIND} in each snapshot. \citet{Yang2018} match the SDSS DR7 galaxies in the above survey area to the ELUCID subhaloes at $z$=0 with a novel neighbourhood abundance matching technique, which we refer to as the SDSS-ELUCID matching catalogue in the following. This approach is similar to the traditional subhalo abundance matching (SHAM, \citealt{Conroy2006, Behroozi2010, Moster2010, Reddick2013, Guo2016}) that links the galaxies and subhaloes through their luminosity (or stellar mass) and subhalo mass (or circular velocity). In addition to this, it takes into account the separation between galaxies and subhaloes and prefers to match the galaxy to the subhalo of appropriate mass in the neighbourhood. As a result, 296,488 galaxies out of 396,069 are assigned to central subhaloes as central galaxies, and 99,581 are assigned to satellite subhaloes as satellite galaxies. We refer the reader to \citet{Yang2018} for more details regarding the neighbourhood abundance matching method and the SDSS-ELUCID matching catalogue. The ELUCID simulation and SDSS-ELUCID matched catalogue are available on the ELUCID website \footnote{\url{https://gax.sjtu.edu.cn/data/ELUCID.html}}.

We investigate the connection between galaxy properties and subhalo properties in the SDSS-ELUCID matching catalogue with machine learning. The galaxy properties we mainly focused on are r-band absolute magnitude $M_{r}$ and the $g-r$ colour, with the magnitudes being K-corrected with evolution corrections to $z=0.1$ according to \citet{Blanton2003b} and \citet{Blanton2007}. We also consider derived physical galaxy properties such as stellar mass and specific star formation rate (sSFR). The subhalo properties we focused on are:
\begin{itemize}
    \item[(1)] $M_{\rm sub}$, the subhalo mass, in units of $\hmsun$;
    \item[(2)] $M_{\rm peak}$, the peak value of $M_{\rm sub}$ over the formation history of the subhalo;
    \item[(3)] $M_{\rm acc}$, the value of $M_{\rm sub}$ when the subhalo accretes onto its host ($M_{\rm acc}=0$ for central subhalo);
    \item[(4)] $r_{\rm half}$, the half mass radius of the subhalo; 
    \item[(5)] $V_{\rm max}$, the maximum circular velocity of the subhalo;
    \item[(6)] $V_{\rm peak}$, the peak value of $V_{\rm max}$ over the formation history of the subhalo;
    \item[(7)] $V_{\rm max, acc}$, the value of $V_{\rm max}$ when the subhalo accretes onto its host; 
    \item[(8)] $V_{\rm disp}$, the velocity dispersion of the subhalo; 
    \item[(9)] $z_{\rm vpeak}$, the redshift when $V_{\rm max} (z_{\rm vpeak}) = V_{\rm peak}$;
    \item[(10)] $z_{\rm mpeak}$, the redshift when $M_{\rm sub} (z_{\rm mpeak}) = M_{\rm peak}$;
    \item[(11)] $z_{\rm acc}$, the redshift when the subhalo accretes onto its host halo;
    \item[(12)] $z_{\rm 0.1/0.3/0.5/0.7/0.9}$, the formation redshift of subhalo, defined by the redshift when the subhalo reaches 0.1/0.3/0.5/0.7/0.9 of its peak mass for the first time;
    \item[(13)] $N_{\rm merg}$, the total number of major mergers (defined by a mass ratio of 1/3 between the progenitors) on the main branch of the subhalo merger tree;
    \item[(14)] $z_{\rm first}$, the redshift of the first major merger of the subhalo;
    \item[(15)] $z_{\rm last}$, the redshift of the last major merger of the subhalo;
    \item[(16)] $t_{\rm sat}$, the total time during which the subhalo is a satellite around the central subhalo, in the unit of $\rm Gyr$;
    \item[(17)] $\lambda$, the spin parameter of the subhalo,
\end{itemize}

and the environmental properties included are: 
\begin{itemize}
    \item[(1)] $\delta_{2.1}$, the matter density smoothed by a Gaussian filter with a smoothing scale of $2.1 \hmpc$;
    \item[(2)] $T_{\rm web}$, cosmic web type, classified as one of knot, filament, sheet, and void according to the eigenvalues of the Hessian matrix \citep{Zhang2009, Paranjape2018} calculated with $\delta_{2.1}$.
\end{itemize}

\subsection{SAM and hydrodynamic simulation} 
\label{sec:sam}

To examine the impact of the mismatch between SDSS galaxies and ELUCID subhaloes on our results, we make use of the \citet{Luo2016} SAM implemented on the subhalo merger tree of ELUCID. As an L-Galaxies model \citep{Guo2011,Guo2013,Fu2013}, it accounts for various galaxy formation processes such as gas cooling, star formation, gas stripping, and feedback from AGN and supernova. In comparison with other SAMs, it introduces an analytic approach to trace the evolution of low-mass subhaloes that fall below the mass resolution of the simulation, improving the modeling of satellite quenching and galaxy clustering.

To further assess the impact of the mismatch, We also perform tests with the TNG-300 hydrodynamic simulation \citep{Marinacci2018, Naiman2018, Nelson2018, Nelson2019, Pillepich2018, Springel2018}. This simulation evolves $2500^3$ dark matter particles of mass $5.9 \times 10^7 \hmsun$ and the same number of baryonic particles of mass $1.1 \times 10^7 \hmsun$ in a cubic box with a length of 205 $h^{-1}{\rm Mpc}$ on a side using the \texttt{AREPO} moving-mesh code \citep{Springel2010}. The Planck cosmology \citep{Planck2016} is adopted, with cosmological parameters $\Omega_{\rm m}=0.31$, $\Omega_{\rm b}=0.0486$, $h=0.677$, and $n_s$=0.97, and $\sigma_8=0.816$. The TNG-300 simulation is an updated version of the original Illustris simulation \citep{Vogelsberger2014, Nelson2015}, with improvements on AGN feedback, galactic wind, and magnetic fields. Compared to the original Illustris, the galaxy colour distribution in TNG is found to be more consistent with observation.

We use the subhaloes from TNG-300-dark, which is a dark-matter-only (DMO) counterpart of the full-physics (FP) TNG-300 simulation. We adopt similar subhalo properties as in Section~\ref{sec:simu} calculated from the \texttt{SUBLINK} merger tree of TNG-300-dark, including $M_{\rm sub}$, $M_{\rm peak}$, $M_{\rm max}$, $V_{\rm max}$, $V_{\rm peak}$, $V_{\rm disp}$, $z_{\rm vpeak}$, $z_{\rm mpeak}$, $z_{\rm acc}$, $z_{\rm 0.1}$, $z_{\rm 0.3}$, $z_{\rm 0.5}$, $z_{\rm 0.7}$, $z_{\rm 0.9}$, $N_{\rm merg}$, $z_{\rm first}$, $z_{\rm last}$, $t_{\rm sat}$, $\lambda$. To assign galaxies to DMO subhaloes, We apply the matching catalogue between the subhaloes of the DMO and FP runs in \citet{Rodriguez-Gomez2015}. In the case that multiple galaxies are matched to one subhalo, we assign the most massive galaxy to the subhalo. To reduce matching noise, we exclude outliers with $\lvert {\rm log}M_{\rm sub,DMO}-{\rm log}M_{\rm sub,FP} \rvert>1$ for the galaxies. The TNG snapshot data, group catalogue, and \texttt{SUBFIND} catalogue are all available on the TNG website \footnote{\url{https://www.tng-project.org/}}.

\subsection{Random forest} 
\label{sec:RF}
We focus on reproducing galaxy properties based on subhalo properties with machine learning techniques to better understand the connection between the two. To accomplish this, we utilise the random forest (RF) model \citep{Breiman2001}, which is highly efficient in capturing complex multi-variate dependencies between input and output variables. The RF model is widely used in galaxy formation studies and shows promising results in reproducing galaxy properties based on halo or subhalo properties \citep{Kamdar2016a, Agarwal2018, Xu2021b, Xu2022}. 

RF is an ensemble of decision trees \citep{Breiman1984} which are constructed by splitting training data into hierarchical nodes. At each node, the training data including feature variables and the target variable is split into lower-level nodes in a way that minimises the cost function (e.g. the Gini impurity for classification tree and mean squared error for regression tree), until the specified maximum level of tree is reached, or the minimum number in node is reached. The predicted output is then calculated from the bottom level of nodes, also known as leaves. For a classification tree, the output is the majority of the target variable of data in the leaf, and for a regression tree, the output is the mean of the target variable of the data in the leaf. Once trained, the RF can be tested using a test sample, and the prediction performance can be estimated by performance scores such as $F_1$ for classification and $R^2$ for regression. To predict galaxy properties, we employ the regression RF in the \texttt{sklearn} package of Python and the $R^2$ score. For all the RF analyses in this work, we use 60\% of the original data as training sample and the rest as test sample. 

\section{Predicting SDSS galaxy properties} 
\label{sec:sdsspred}
We construct RF models for predicting galaxy $r$-band absolute magnitude $M_r$ and $g-r$ colour separately. These models are trained using galaxies selected from the SDSS-ELUCID matching catalogue, and the subhalo properties listed in Section~\ref{sec:simu} are used as input variables. With the predicted $M_r$, we compare the luminosity function and galaxy-matter cross-correlation in different $M_r$ bins to those in observation. We then compare the predicted colour distribution to that of the SDSS.

\subsection{Subhalo mass completed sample} 
\label{sec:zlim}
For training the RF, we first select an appropriate sample from the SDSS-ELUCID matching catalogue. Since only the galaxies brighter than a specific magnitude threshold can be observed at fixed redshift, the low-mass subhaloes with faint galaxies are likely underrepresented in the SDSS-ELUCID matching catalogue, which is also known as the Malmquist bias. With this bias, the number density of subhaloes of fixed mass decreases beyond a certain redshift, which we refer to as the limited redshift $z_{\rm lim}$. In other words, the subhalo population of this mass is incomplete above $z_{\rm lim}$. For a specific low subhalo mass, galaxies residing in early-formed subhaloes with luminosities higher than average are more likely to be observed, while those with luminosities lower than average may fall below the detection limit of the survey. This leads to a biased luminosity-subhalo mass relation for the low-mass subhaloes in the SDSS-ELUCID catalogue. If the RF captures this biased relation, the predicted magnitude would be brighter than expected at fixed subhalo mass. It will also introduce biases in the relationships between other galaxy properties and subhaloes since the early-formed subhaloes are more represented in observation. To avoid this kind of bias, it is necessary to select the subhaloes with redshift smaller than their $z_{\rm lim}$. 

\begin{figure}
    \centering
    \begin{subfigure}[h]{0.48\textwidth}
    \includegraphics[width=\textwidth]{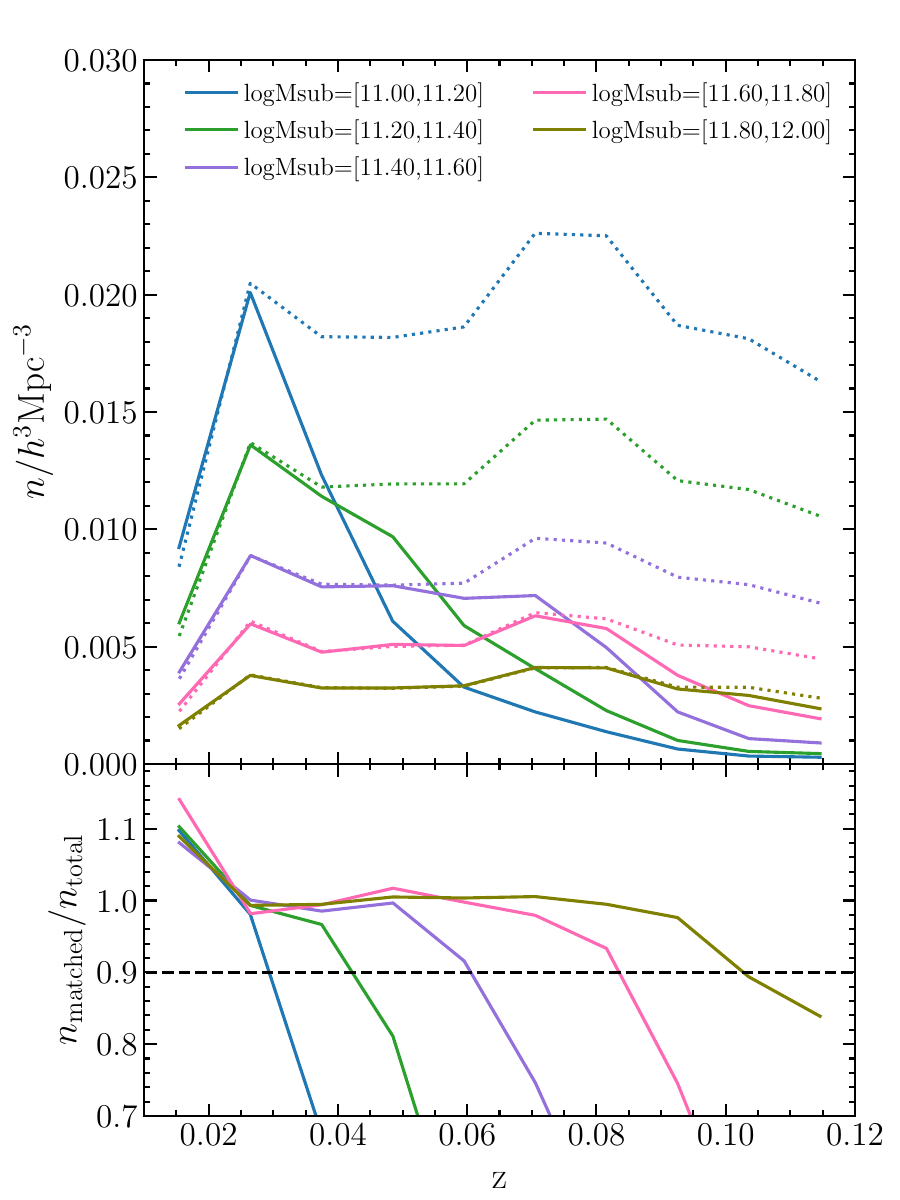}
    \end{subfigure}	    
    \hfill
\caption{Top: subhalo number density as a function of redshift at fixed ${\rm log} M_{\rm sub}$ for SDSS-ELUCID matched subhaloes (solid) and ELUCID subhaloes in the SDSS region (dotted). A few selected ${\rm log} M_{\rm sub}$ bins are shown with different colours. Bottom: the ratio between the number density of SDSS-ELUCID matched subhaloes and ELUCID subhaloes in the SDSS region. The complete threshold of 0.9 is indicated by the black dashed line.}
\label{fig:z-msublim}
\end{figure}

In Figure~\ref{fig:z-msublim}, we compare the number densities of SDSS-ELUCID matched subhaloes (solid) to all the ELUCID subhaloes (dashed) in the SDSS region in ${\rm log} M_{\rm sub}$ bins of 0.2 dex as a function of redshift. Different colours represent ${\rm log} M_{\rm sub}$ bins in the range of [11, 12]. The total number densities of SDSS region subhaloes are approximately constant across the redshift range, except for a bump near z=0.08, which may be caused by the well-known Sloan "great wall" structure. In contrast, the number densities of SDSS-matched subhaloes deviate from those of the SDSS region subhaloes and decline beyond specific redshifts, which increase with subhalo mass. This indicates again that the subhalo sample matched to SDSS galaxies may be incomplete due to the Malmquist bias. The impact of Malmquist bias vanishes for subhaloes of ${\rm log} M_{\rm sub}>12$. The bottom panel shows the ratio between the two number densities $n_{\rm matched}/n_{\rm total}$. For each subhalo mass bin, we define the limited redshift $z_{\rm lim}$ at which the ratio drops to 0.9 (shown by the dashed line). By interpolating between the mass bins, a $z_{\rm lim}$ can be calculated for each galaxy (subhalo) in the SDSS-ELUCID matched sample according to the subhalo mass. We then select the galaxies with redshift below their $z_{\rm lim}$. As a result, 201,980 galaxies are selected from the original 396,069 galaxies for our RF analysis. We refer to this sample as the $z_{\rm lim}$-selected sample. We also perform a test calculating $z_{\rm lim}$ with ${\rm log} M_{\rm peak}$ instead of ${\rm log} M_{\rm sub}$, and the result is similar.

\subsection{r-band magnitude} 
\label{sec:mrpred}

\begin{figure*}
    \centering
    \begin{subfigure}[h]{0.9\textwidth}
    \includegraphics[width=\textwidth]{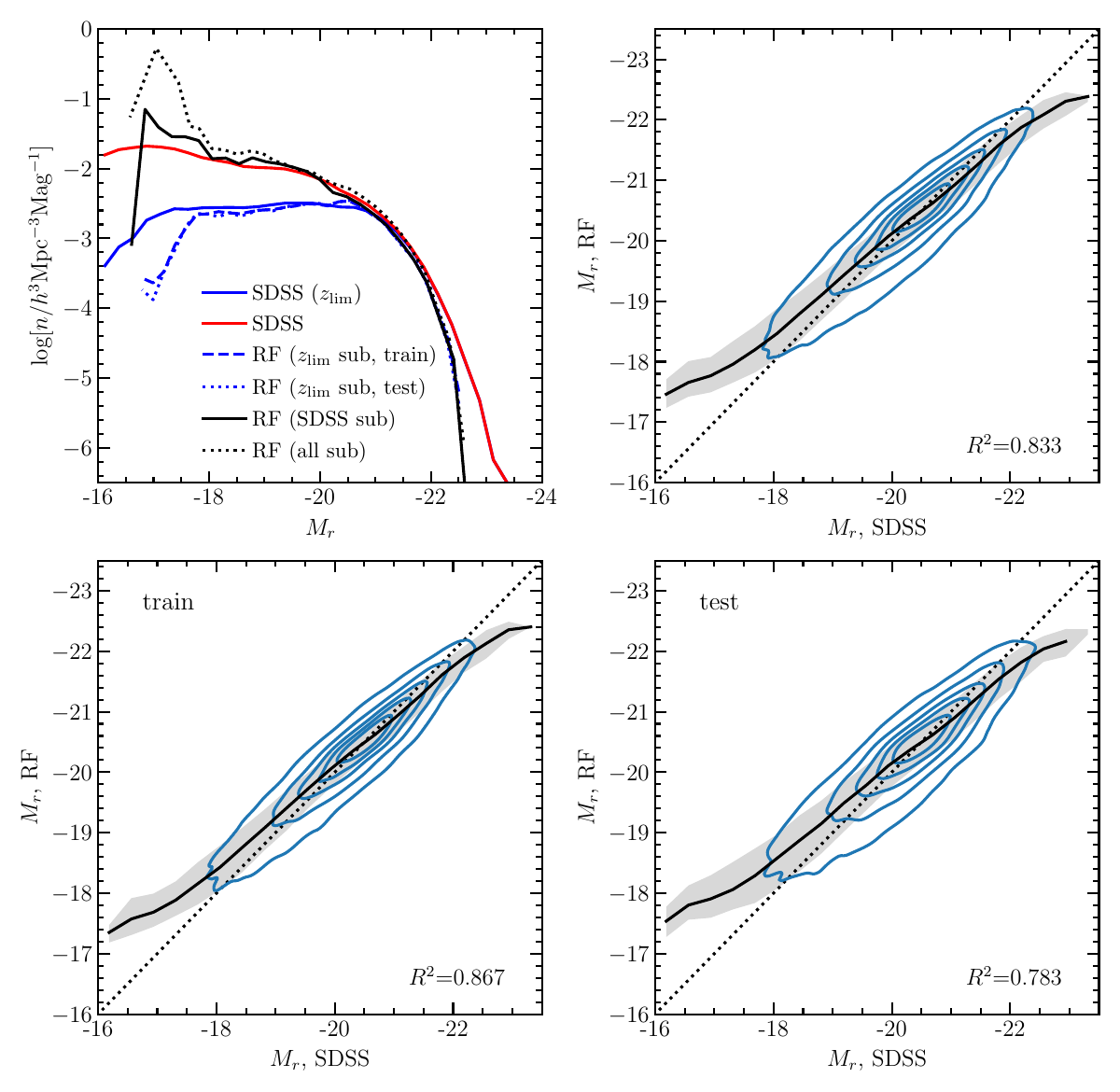}
    \end{subfigure}
\caption{$M_r$ Prediction trained on the $z_{\rm lim}$-selected SDSS-ELUCID catalogue. Top-left: luminosity function of the $z_{\rm lim}$-selected SDSS galaxies (solid blue) and RF prediction separated into training sample (blue dashed) and test sample (blue dotted). The solid black curve shows the predicted $M_r$ applying the trained RF on all subhaloes in the SDSS-ELUCID sample, and the solid red shows the measurement of galaxies in the same sample. The dotted black indicates the RF prediction on all ELUCID subhaloes. Top-right: distribution of comparison between SDSS $M_r$ (x-axis) and predicted $M_r$ (y-axis) of the $z_{\rm lim}$-selected sample, shown by the blue contours (20\%, 40\%, 60\%, 80\%, 95\% of the sample). The black solid and shadow indicate the median and 16\%-84\% of the prediction at fixed SDSS $M_r$. Equality is shown by the black dashed line along the diagonal direction. Bottom-left/right: comparison between SDSS and prediction in the training/test sample.}
\label{fig:sdss-mr}
\end{figure*}

The results of the $M_r$ prediction are shown in Figure~\ref{fig:sdss-mr}. In the top-left panel, we compare the luminosity function (LF) of the SDSS $M_r$ (blue solid) of the $z_{\rm lim}$-selected sample with the corresponding RF predictions (blue dashed for training and blue dotted for test set). To measure the LF, we adopt the $V_{\rm max}$ method, which determines the maximum volume in which the galaxy can be observed above the flux limit of the survey (note that this is different from the subhalo maximum circular velocity $V_{\rm max}$). For each galaxy, a weight of inverse $V_{\rm max}$ is assigned for number counting. The RF predictions demonstrate good agreement with the SDSS measurement within the magnitude range of $-22<M_r<-18$. However, discrepancies arise at both the bright end ($M_r<-22$) and the faint end ($M_r>-18$), where the prediction is lower than the SDSS. This is not surprising, as the machine learning methods are unable to reproduce 100\% variance of the input data and tend to underpredict extreme values \citep{Agarwal2018}. The RF predictions on the training and test sample are in excellent agreement, indicating that the construction of the RF is appropriate and the prediction result is reliable. 

We then apply this trained RF to all subhaloes in the SDSS-ELUCID sample and show the predicted $M_r$ LF by the black solid. For comparison, the direct measurement of the SDSS LF of the same sample is shown by the red solid. Similar to the result of the $z_{\rm lim}$-selected sample, the prediction is consistent with the direct measurement within the range of $-22<M_r<-18$. As the SDSS region only covers a fraction of the ELUCID volume, we also apply the trained RF model to all the subhaloes in the ELUCID simulation and show the predicted LF with the black dotted curve. It is again very similar to the SDSS measurement, with the exception of the bright and faint ends.  A bump exhibits at $M_r>-18$, which is likely attributed to the low abundance of faint galaxies hosted by low-mass subhaloes in our training sample. In addition to this, the cosmic variances may also contribute to the discrepancy at the faint end. As highlighted in \citet{Chen2019}, the faint end slope of the LF was significantly underestimated due to the cosmic variances in the SDSS observation.

The top-right panel presents a direct comparison between the SDSS $M_r$ (x-axis) and the predicted $M_r$ (y-axis) for all galaxies in the $z_{\rm lim}$-selected sample. The blue contours show the 20\%, 40\%, 60\%, 80\%, 95\% of the data distribution, and the black solid (shadow) shows the median (16\%-84\%) of predicted $M_r$ at fixed SDSS $M_r$. The black dashed line along the diagonal indicates equality between the prediction and SDSS values. Overall, the prediction is consistent with SDSS along the equality except for the faint and bright end. For galaxies fainter than $M_r \sim -20$, the RF tends to predict brighter magnitudes, while the trend is reversed for galaxies brighter than $M_r \sim -20$. Scatters exist in the prediction at fixed SDSS $M_r$, with smaller scatter for brighter galaxies compared to fainter ones. To quantify the performance of the prediction, we provide the $R^2$ score which describes the fraction of the variance in the target variable (e.g. $M_r$ in this case) captured by the prediction at the bottom right of the panel. As $R^2=1$ represents a perfect prediction that recovers the full variance in the target variable, an $R^2$ of 0.8 indicates that our prediction captures a significant fraction of the variance in SDSS $M_r$. 

The bottom-left and bottom-right panels show the same comparison for the training sample and test sample, respectively. The $R^2$ of the training sample is slightly higher than that of the full sample, and the $R^2$ of the test sample is slightly lower. This is reasonable since the RF is data-driven, and the model is trained to fit the training sample with a priority. We also perform the same analysis to predict the stellar mass, and the result (shown in Appendix~\ref{sec:appendix}) is very similar to that of the $M_r$ prediction.

\begin{figure*}
    \centering
    \begin{subfigure}[h]{0.99\textwidth}
    \includegraphics[width=\textwidth]{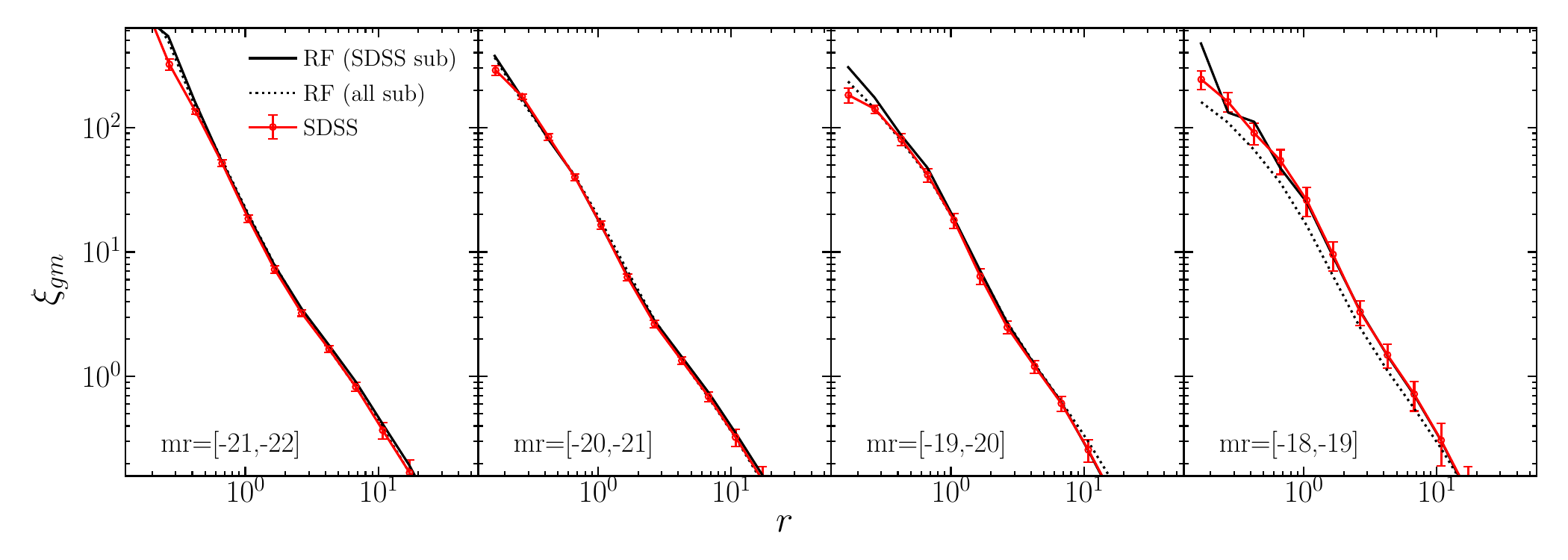}
    \end{subfigure}
\caption{Galaxy-matter cross-correlation of SDSS $M_r$ samples and the predictions. Four $M_r$ bins are shown in four panels. In each panel, the original SDSS cross-correlation is shown by the red solid, and the error bars are measured from 16 jackknife samples. The cross-correlation of predicted $M_r$ of SDSS subhaloes (all subhaloes) is shown by the black dashed (dotted). }
\label{fig:sdss-xic}
\end{figure*}

We then proceed to compare the $M_r$-dependent galaxy clustering in SDSS and the predictions. To measure the SDSS clustering, we construct four volume-limited $M_r$ bin samples in which the sample completeness is ensured.  In other words, the apparent magnitudes of all galaxies in each bin fall in the detection limits of the survey from $m_r$=14.5 to $m_r$=17.72 \citep{Zehavi2005}. To obtain a higher signal-to-noise signal, we calculate the two-point galaxy-matter cross-correlation using the estimator $\xi_{\rm gm}$=DD/DR-1 in the ELUCID coordinate instead of the galaxy-galaxy auto-correlation, where DD is the number of galaxy-matter pairs, and DR is the number of galaxy-random pairs. The positions of subhaloes serve as the positions of their matched galaxies. 

The SDSS clustering of each $M_r$ bin is illustrated by the red solid curve in each panel of Figure~\ref{fig:sdss-xic}. The black solid curve shows the prediction of SDSS-matched subhaloes, and the black dotted curve indicates the prediction from all subhaloes in ELUCID. In the three bright bins where $-22<M_r<-19$, both the prediction of SDSS-matched subhaloes and all subhaloes consist with the SDSS, except for very small scales. It is worth noting that for the clustering of SDSS-matched prediction, we still utilise the position of subhaloes, so the clustering discrepancy is solely due to the prediction of $M_r$. 

In the faintest bin where $-19<M_r<-18$, the prediction of the SDSS-matched sample still agrees with SDSS measurement. However, the prediction from all subhaloes in ELUCID exhibits a lower clustering amplitude than SDSS on all scales. This discrepancy can be attributed to the bump of the black dotted curve in Figure~\ref{fig:sdss-mr}, which could be a result of the scarcity of low-mass subhaloes in the training sample and therefore the low accuracy of $M_r$ prediction in these subhaloes. 

\subsection{g-r colour} 
\label{sec:colorpred}

\begin{figure*}
    \centering
    \begin{subfigure}[h]{0.99\textwidth}
    \includegraphics[width=\textwidth]{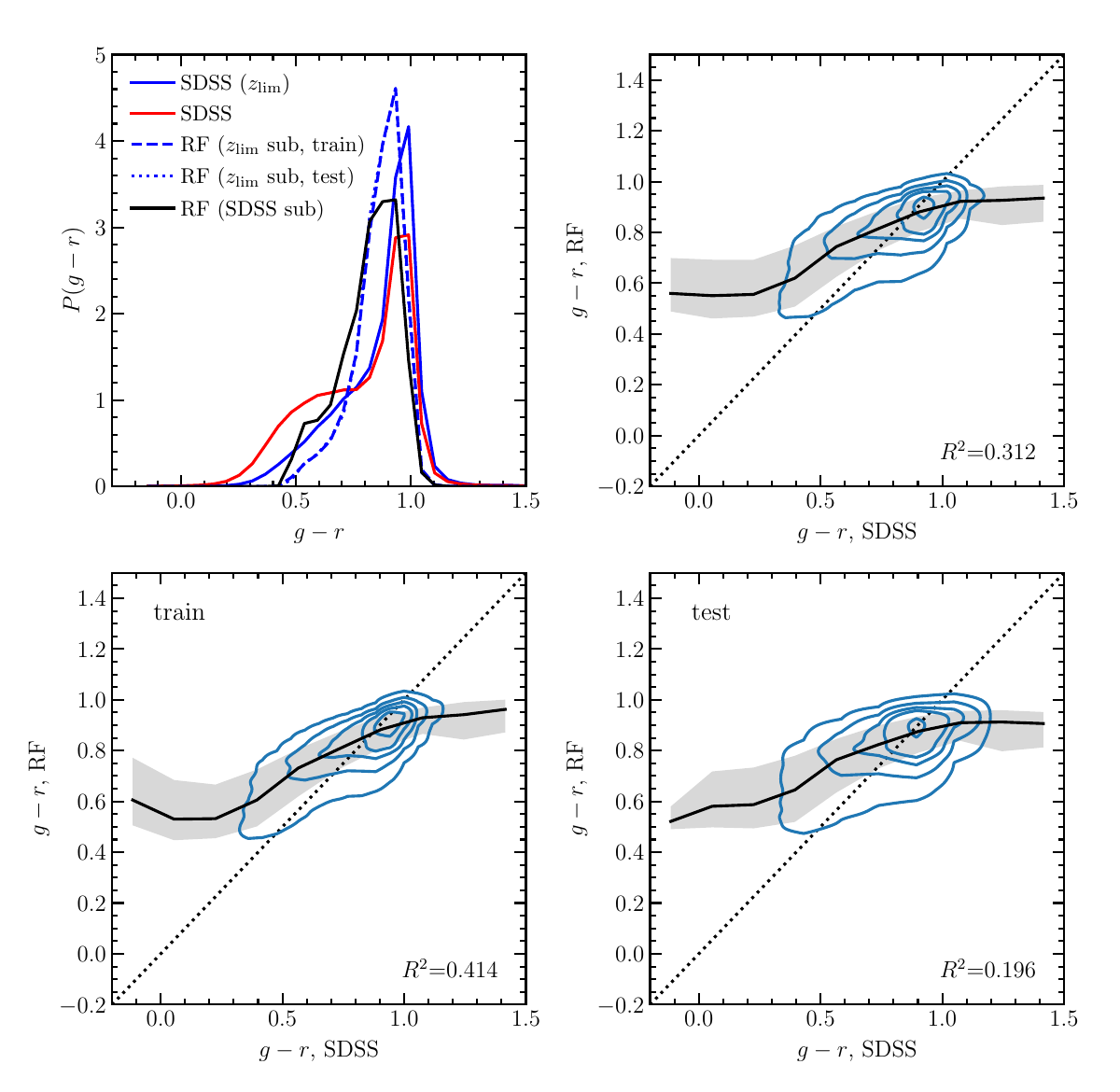}
    \end{subfigure}
\caption{$g-r$ colour prediction trained on the $z_{\rm lim}$-selected SDSS-ELUCID catalogue. Top-left: SDSS colour distribution of the $z_{\rm lim}$ sample (blue solid) and the RF predictions of the training (blue dashed) and test (blue dotted) sets from this sample. The application of this RF to all SDSS-ELUCID subhaloes is shown by the black solid, and the corresponding true SDSS colour distribution of these subhaloes is shown by the red solid. Top-right: comparison between SDSS colour (x-axis) and predicted colour (y-axis). Bottom-left/right: comparison between SDSS and prediction in the training/test sample.}
\label{fig:sdss-color}
\end{figure*}

In addition to the $M_r$, we also train the RF model to predict $g-r$ colour with the subhalo properties, and the results are shown in Figure~\ref{fig:sdss-color}. The top-left panel displays the distribution of the SDSS colour of the $z_{\rm lim}$-selected sample (blue solid) and the RF prediction separated into training (blue dashed) and test sets (blue dotted, overlapping with the blue dashed). We then apply this RF on all subhaloes of the SDSS-ELUCID catalogue and show the prediction by the black solid, and also provide the SDSS colour distribution of the same subhaloes by red solid for comparison. The SDSS colour distribution consists of a narrow red peak around $g-r=1$ and a smooth blue component in the range of $0.4<g-r<0.7$, and only the red peak remains after the $z_{\rm lim}$ selection. However, the red peak of the RF prediction shifts towards lower values of $g-r$. Additionally, the width of the predicted distribution is narrower than that of the SDSS, indicating that extreme red and blue values are not fully recovered by the RF. Since the RF is trained solely on the red peak galaxies, it is not able to recover the blue component when applied to all subhaloes in the SDSS-ELUCID matched catalogue.  

In the top-right panel, we show the comparison between the SDSS colour (x-axis) and the predicted colour (y-axis). The overall trend deviates more noticeably from the diagonal compared to that of the $M_r$ prediction, and the $R^2$ score ($\sim 0.3$) is significantly lower. The bottom-left and bottom-right panels display the prediction for the training sample and test sample, respectively. The $R^2$ of the training (test) sample is slightly higher (lower) than that of the full sample but still indicates a similar level of prediction accuracy. Instead of using the inferred assembly properties characterising subhalo formation history such as $z_{\rm 0.1/0.3/0.5/0.7/0.9}$, we also input the original merger tree information to the RF by using the subhalo masses of 21 snapshots from $z=4.86$ to $z=0$, and masses are set to zero if the subhaloes are not identified in early redshifts. The result is very similar to that using the inferred assembly properties. We also build RF models for central and satellite galaxies separately, but no significant improvements in the prediction are found. This indicates that predicting the SDSS galaxy colour with subhalo properties is more challenging than predicting $M_r$. We also train the RF to predict the SFR and specific sSFR of SDSS galaxies based on subhalo properties. The results are shown in Appendix~\ref{sec:appendix}. The $R^2$ of sSFR prediction is similar to that of the colour, while the $R^2$ of SFR is much lower.

The reasons for the low-accuracy colour prediction are complicated. Firstly, the correlation between galaxy colour and subhalo properties may be weak in SDSS, and baryonic processes such as AGN feedback could have more significant effects on galaxy colour. Secondly, noise in the galaxy-subhalo relation of the training sample may raise from possible mismatches between the SDSS galaxies and ELUCID subhaloes. It is difficult to test the first possibility directly since subhalo or halo properties such as formation redshift are difficult to measure in observation. Empirical models can be used to infer the correlation between galaxy colour and halo property. For example, \citet{Hearin2013} propose an age-matching model that assumes a monotonic relation between galaxy colour and subhalo assembly property to reproduce colour-dependent galaxy clustering. On the other hand, \citet{XuHaojie2018} propose a conditional colour-magnitude distribution model that assumes magnitude and colour depend purely on halo mass and find that it can also reproduce the observed galaxy clustering dependence on colour reasonably well. In these two models, the former suggests a non-zero relation between colour and halo assembly history, while the latter suggests an independent trend. This indicates that the conclusion can be model-dependent, and further investigations are needed to resolve this debate. 

In this study, we will focus on investigating the second possible reason mentioned above, which is the mismatch between SDSS galaxies and ELUCID subhaloes. It is important to note that the term "mismatch" here refers not only to errors in matching caused by the neighbourhood abundance matching method, but also to other sources of noise that could introduce biases in the galaxy-subhalo relation. All the RF studies above are based on the assumption that the matching is accurate, or in other words, that the true subhalo properties of a galaxy can be accurately recovered by those of the matched ELUCID subhalo. However, this is not guaranteed, especially for the low-mass subhaloes that are expected to host faint galaxies, as these may not be recovered by the constrained simulation. The reconstruction of the matter density from the group catalogue only uses groups of mass above ${\rm log} M_{\rm group}=12$ and applies a Gaussian kernel with a smoothing scale of 2 $\hmpc$ \citep{Wang2016}. As a result, information on haloes and subhaloes below this mass scale and length scale is lost, and the reconstructed (sub)haloes could differ from the actual ones. Matching galaxies to these subhaloes could introduce noises to the galaxy-halo relations compared to the true ones. Therefore, it is necessary to consider the mismatch effect when analysing the galaxy-halo relations based on this galaxy-subhalo matching catalogue. In the following section, we will perform tests to investigate the impact of the mismatch effect on our RF results using SAM and hydrodynamic simulation.
 
\section{Mismatch effect in prediction} 
\label{sec:mismatch}

\subsection{Mismatch effect using SAM} 
\label{sec:missam}

\begin{figure*}
    \centering
    \begin{subfigure}[h]{0.99\textwidth}
    \includegraphics[width=\textwidth]{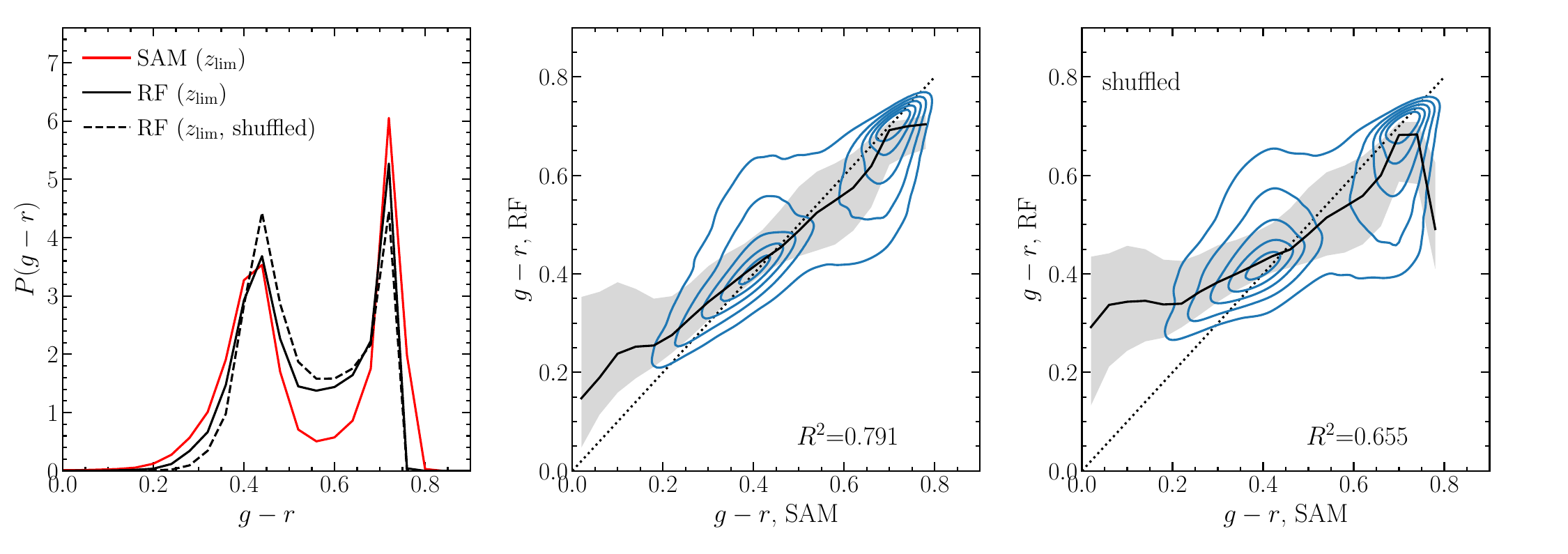}
    \end{subfigure}
\caption{$g-r$ colour prediction based on original SAM and shuffled SAM. Left: SAM colour distribution of SDSS $z_{\rm lim}$-selected subhaloes (solid red), predicted colour of these subhaloes (black solid), and the prediction based on the shuffled sample (black dashed). Middle: comparison of SAM colour and predicted colour for SDSS-matched subhaloes. Right: comparison of SAM colour and prediction based on the shuffled sample for these subhaloes.}
\label{fig:sam-color}
\end{figure*}

In this section, we aim to test the potential impact of the mismatch effect between SDSS galaxies and ELUCID subhaloes by creating a similar mismatch in galaxies of a SAM model implemented on ELUCID \citep{Luo2016}. Since we regard the mismatch effect as noise in the galaxy-subhalo relation, we mimic it by randomly shuffling the SAM galaxies in subhaloes within narrow $M_{\rm peak}$ bins of 0.2 dex in the vicinity of $5\hmpc$ cubic cells. The constrain of narrow $M_{\rm peak}$ bin maintains a relatively reasonable stellar mass-$M_{\rm peak}$ relation, consistent with the principle of neighbourhood abundance matching when assigning SDSS galaxies to ELUCID subhaloes. Shuffling in the neighbourhood of $5\hmpc$ cells is in line with the advantage of the constrained simulation that it can recover the subhalo distribution at this scale. For example, \citet{Yang2018} investigate the separation between galaxy and subhalo pairs in the SDSS-ELUCID matched catalogue and find that most of the pairs are separated below $\sim 5\hmpc$ in both $r_{p}$ and $\pi$ directions. The shuffling breaks the original connection between galaxy properties and subhalo properties other than $M_{\rm peak}$, thus adding noise to the true galaxy-subhalo relation. 

We subsequently construct RF models to predict galaxy colour using the original SAM galaxy-subhalo pairs and the shuffled pairs, respectively. To access a reasonable estimation of the mismatch effect, we use the SAM galaxies hosted by the subhaloes of the $z_{\rm lim}$-selected sample before shuffling. The left panel of Figure~\ref{fig:sam-color} shows the SAM colour distributions of galaxies in $z_{\rm lim}$-selected subhaloes (red solid) and the RF prediction (black solid), which are highly similar. The middle panel displays the two-dimensional distribution. Generally, the contours are aligned with equality with small deviation. The black solid with shadow indicates the median and 16\%-84\% of prediction at fixed SAM colour bins. The large deviation at $g-r<0.2$ of SAM is possibly due to the low number of extreme blue galaxies in this range. The $R^2$ of the prediction is $\sim 0.8$, significantly higher than that in the SDSS prediction. This indicates that the galaxy-subhalo connection in the SAM is much stronger, which is consistent with the construction of the SAM. \citet{Xu2022} find that adding galaxy properties such as black hole mass and cold gas mass can further improve the prediction of the SAM colour. 

Recently, \citet{Jespersen2022} propose a graph neural network method to predict several SAM galaxy properties based on halo merger trees. Unlike traditional machine learning methods where the input features are halo properties extracted from the merger tree, their model uses the merger tree itself as input, maximizing the information obtained from the growth history of the halo. Their prediction performance of SFR is impressively higher ($R^2=0.876$) compared to previous studies in the literature. We also perform a test predicting the SFR with RF and find that the $R^2$ score is 0.864, which is very similar to that of the graph neural network. This indicates that the RF is capable of capturing the connections between galaxy properties and halo or subhalo properties if exist.   

Back to the left panel of Figure~\ref{fig:sam-color}, the black dashed curve indicates the prediction based on the shuffled sample. The prediction still features the blue and red peaks, but the red peak is slightly lower than that in the original SAM, and the blue peak is slightly higher. The overall recovered colour range is narrower, and some of the extreme blue and red values are missing compared to the prediction before shuffling. The right panel is the two-dimensional comparison between the shuffled prediction and the original SAM. The deviation from equality is larger than that in the middle panel, especially at $g-r<0.4$, and the scatter in the prediction at fixed SAM colour is also larger. The $R^2$ value of 0.655 is lower than that before shuffling. 

With the noises in the galaxy-subhalo relation introduced by the shuffling, the performance of RF colour predicting is impacted. However, even with shuffling, the $R^2$ score of the prediction is still higher than that of the SDSS prediction. We find from the RF that the most important subhalo feature for predicting SAM colour is $V_{\rm peak}$, which is highly correlated with $M_{\rm peak}$ and likely remains similar after shuffling. Other relatively important subhalo features for the prediction are subhalo assembly properties such as $z_{\rm acc}$ and $z_{\rm 0.1/0.3/0.5/0.7/0.9}$. Although the shuffling process reassigned these subhalo properties for a given galaxy, the correlations between galaxy and subhalo properties may not be completely removed due to the constraints of the shuffling. This will be further demonstrated by the correlation coefficients before and after shuffling in Section~\ref{sec:comp}. As a result, the galaxy colour can still be partially reproduced after the shuffling. If the colour-subhalo relation in the real Universe is similar to that in the SAM, the RF could capture this relation with an $R^2$ of approximately 0.6, accounting for possible mismatches. Thus, it is likely that the connection between colour and subhalo properties in the real Universe is not as strong as that in the SAM. As a further step, we perform a similar test using the TNG300 hydrodynamic simulation and compare it with the SDSS prediction in the following section.

\subsection{Mismatch effect using TNG300} 
\label{sec:mistng}

\begin{figure}
    \centering
    \begin{subfigure}[h]{0.48\textwidth}
    \includegraphics[width=\textwidth]{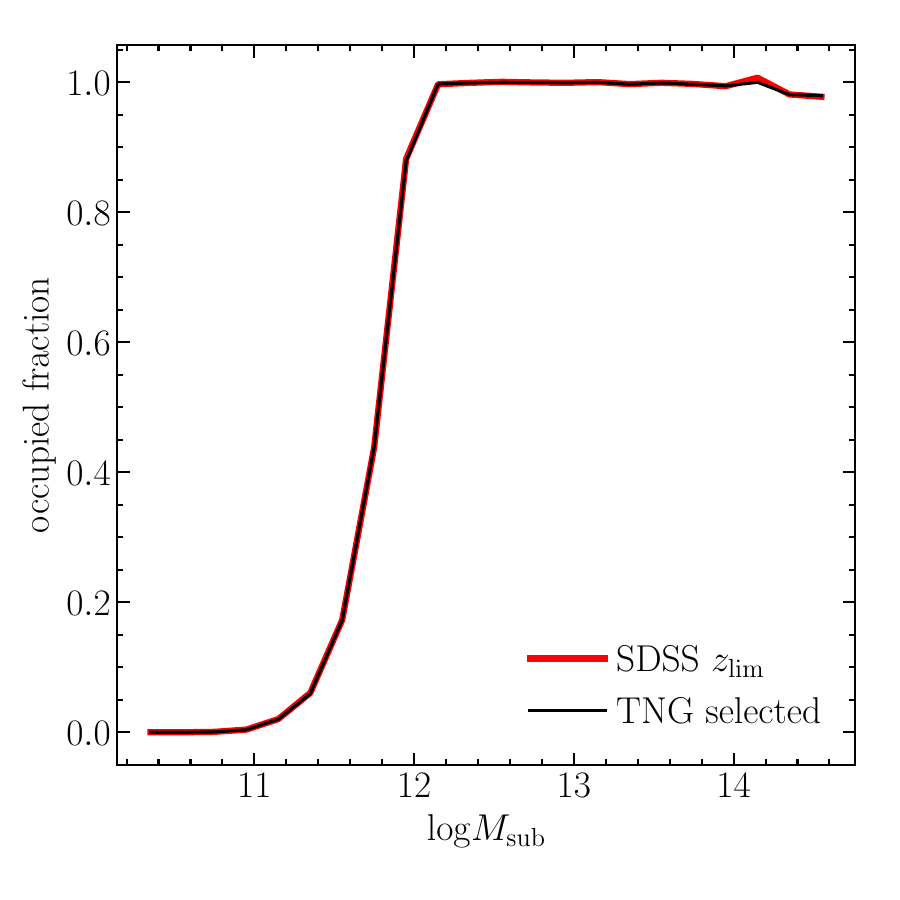}
    \end{subfigure}	    
    \hfill
\caption{The fraction of subhaloes occupied by SDSS $z_{\rm lim}$-selected galaxies as a function of ${\rm log} M_{\rm sub}$. Red solid indicates the occupied fraction in the SDSS $z_{\rm lim}$ sample, which is defined as the ratio of the number of subhaloes in this sample to the total number of subhaloes in the SDSS region of ELUCID. The black solid indicates the occupied fraction in the selected sample of TNG300.}
\label{fig:tng-emtpyfrac}
\end{figure}

Without the SDSS region, comparisons between the TNG300 predictions and the SDSS or SAM predictions are indirect. Since the SDSS galaxies are matched to a fraction of ELUCID subhaloes in the corresponding SDSS region, and we select SDSS galaxies according to $z_{\rm lim}$ to ensure the completeness of subhaloes, some subhaloes in the SDSS region of ELUCID are empty (i.e. not occupied by $z_{\rm lim}$-selected SDSS galaxies). Since more massive subhaloes tend to host brighter galaxies that are more likely to be observed, the occupied fraction of subhaloes will increase with ${\rm log} M_{\rm sub}$. To account for this effect in TNG300, we measure the occupied fraction as a function of ${\rm log} M_{\rm sub}$ in the SDSS $z_{\rm lim}$ catalogue and select a random sample of galaxies in TNG300 which can reproduce this trend.  

Figure~\ref{fig:tng-emtpyfrac} displays the occupied fraction in both the SDSS $z_{\rm lim}$ sample (red solid) and the selected TNG300 sample (black solid). The occupied fraction is $\sim 0$ for ${\rm log} M_{\rm sub}<11$ and rapidly increases to $\sim 1$ at ${\rm log} M_{\rm sub}\sim 12$. This implies that the abundance of low-mass subhaloes is largely suppressed in observation, while the subhaloes of ${\rm log} M_{\rm sub}>12$ are barely affected. The advantage of this selection in TNG300 is that it can create a training sample where the subhalo population is similar to that of the SDSS training sample. This is important because the machine learning performance of colour prediction might depend on ${\rm log} M_{\rm sub}$.

\begin{figure*}
    \centering
    \begin{subfigure}[h]{0.99\textwidth}
    \includegraphics[width=\textwidth]{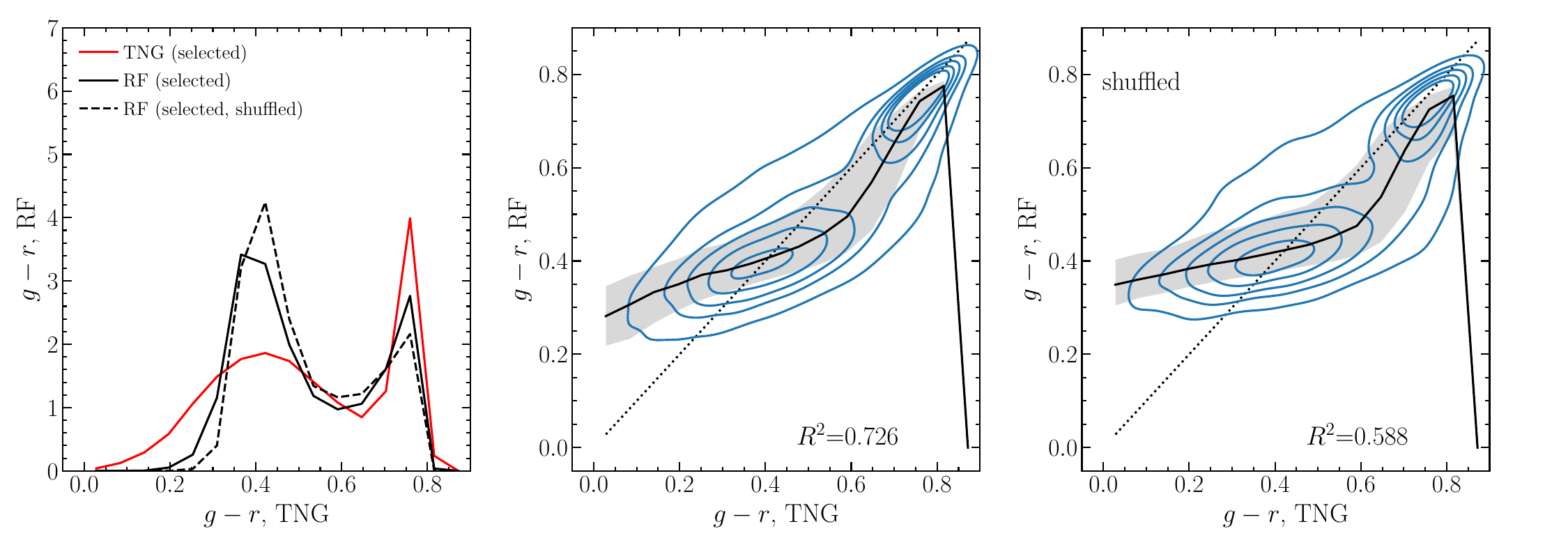}
    \end{subfigure}
\caption{$g-r$ colour Prediction based on original TNG and shuffled TNG. Left: TNG colour distribution of selected subhaloes (red solid), prediction of these subhaloes (black solid), and prediction based on the shuffled sample (black dashed). Right: comparison of the TNG colour and the prediction.  Right: comparison of TNG colour and prediction based on the shuffled sample.}
\label{fig:tng-color}
\end{figure*}

To investigate the effect of mismatch on RF colour prediction using the TNG300 simulation, we implement the shuffling strategy described in Section~\ref{sec:missam} on the selected sample, shuffling galaxies in subhaloes of fixed subhalo mass bins (0.2 dex) in cells of $5\hmpc$. Similar to the SAM, we construct RF models to predict galaxy colour with subhalo properties before and after shuffling and present the results in Figure~\ref{fig:tng-color}. The left panel shows the colour distribution of the selected TNG sample (red solid) and the corresponding prediction (black solid). The TNG colour distribution shows a narrow red peak at $g-r=0.75$ and a broad blue peak around $g-r=0.4$. The prediction successfully captures the red peak, but the predicted blue peak is narrower and higher than that in TNG, and the amount of extreme blue galaxies with $g-r<0.3$ are underestimated. In the middle panel, the deviation of the prediction is mainly seen at $g-r<0.4$ where the predicted values are higher, and the prediction at the red end aligns more closely with TNG. The $R^2$ score for the prediction is 0.726, which is comparable to that in the SAM. However, the performance of TNG RF in recovering the blue colour is relatively worse than that of the SAM. 

The black dashed in the left panel represents the RF prediction based on the shuffled sample. Compared to the prediction before shuffling, both the predicted red and blue peaks deviate more from those in the original TNG, in the way that the red peak is lower and the blue peak is higher. Moving to the right panel which illustrates the two-dimensional distribution of the shuffled prediction and the original TNG, we find that the deviation from equality is also more pronounced, with an $R^2$ score of 0.588. Compared to the SAM results in Figure~\ref{fig:sam-color}, the mismatch effect shows a similar impact on the RF prediction of the TNG sample. 

It is worth noting that both the $R^2$ of SAM and TNG prediction after shuffling are higher than that of the SDSS prediction. Assuming that the SDSS-ELUCID matched catalogue is also subject to a similar mismatch effect, it is reasonable to infer that the true connection between galaxy colour and subhalo properties in SDSS is weaker than those in the SAM and TNG before shuffling. This suggests that the galaxy colour in the real Universe may also depend on baryonic processes such as AGN feedback. In the next subsection, we will compare the galaxy-subhalo relation in SDSS, SAM, and TNG in more detail in terms of the correlation coefficient between galaxy properties and subhalo properties.

\subsection{Comparison between SDSS, SAM, and TNG} 
\label{sec:comp}

\begin{figure*}
    \centering
    \begin{subfigure}[h]{1.1\textwidth}
    \includegraphics[width=\textwidth]{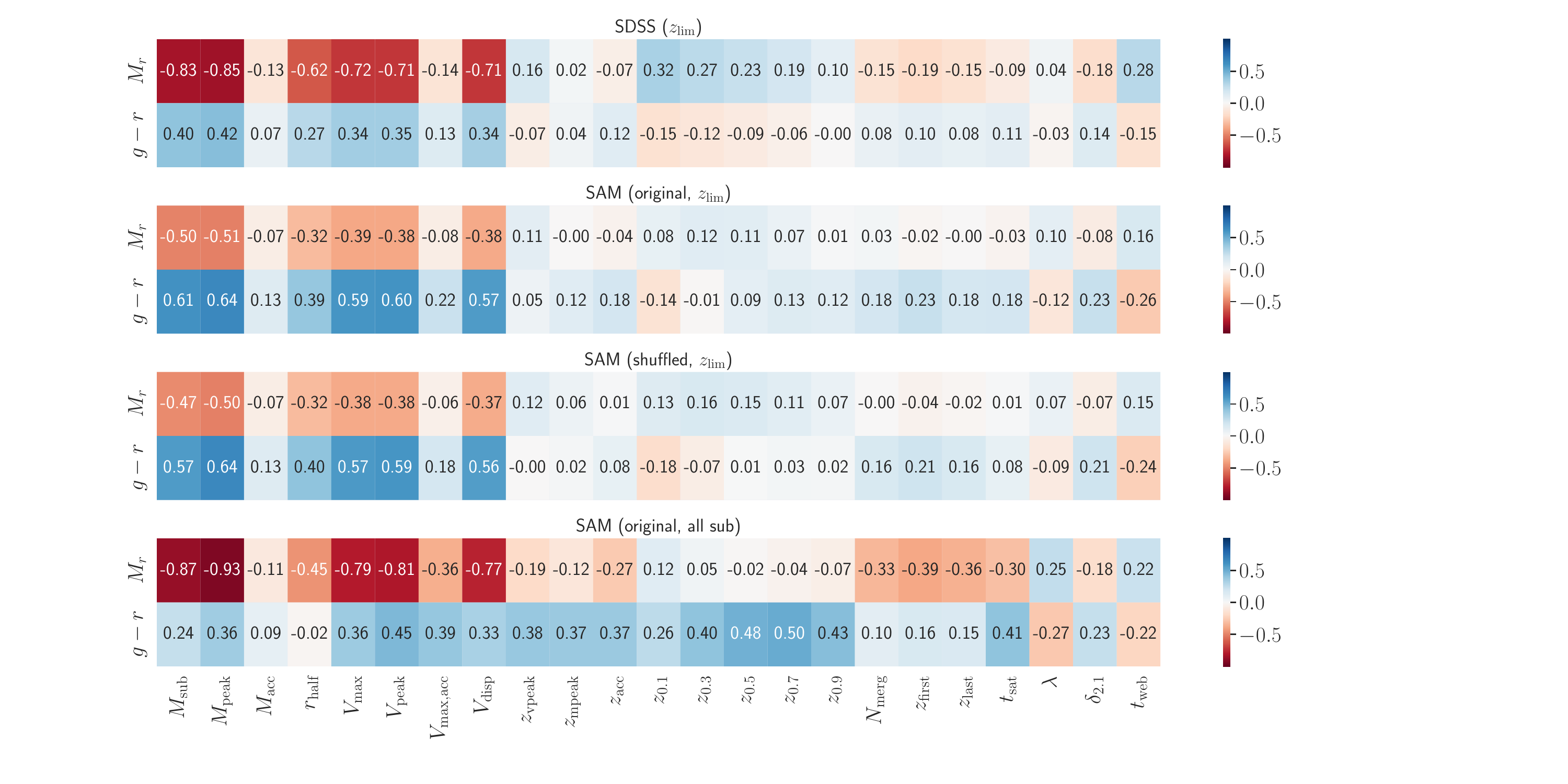}
    \end{subfigure}
\caption{Pearson correlation coefficient between galaxy properties ($y$-axis) and subhalo properties ($x$-axis). From top to bottom, the samples are $z_{\rm lim}$-selected SDSS galaxies (subhaloes), original SAM galaxies of these subhaloes, shuffled SAM, and original SAM of all subhaloes above $M_{\rm sub}$=10.}
\label{fig:corrcoef-sam}
\end{figure*}

To further investigate the differences in the galaxy-subhalo relations between the SDSS, SAM, and TNG samples, we calculate the Pearson correlation coefficient $\rho$ between each pair of galaxy properties and halo properties. The correlation coefficient is a statistical measure that quantifies the strength and direction of the correlation between two variables. It ranges from -1 to 1, and values close to 1 (-1) indicate strong positive (negative) correlations, while values close to 0 indicate weak correlations. In Figure~\ref{fig:corrcoef-sam}, we show the correlation coefficients between SDSS or SAM galaxy properties ($y$-axis) and ELUCID subhalo properties ($x$-axis). Subhaloes with non-physical $z_{\rm 0.1/0.3/0.5/0.7/0.9}$ (i.e. main branch starts with a fraction of peak mass larger than 0.1/0.3/0.5/0.7/0.9) are excluded when measuring the correlation coefficients related to these properties. The colour coding indicates the correlation coefficients, with reddish for positive correlations and blueish for negative correlations. 

The top panel shows the correlation coefficients in the SDSS-ELUCID matched sample with $z_{\rm lim}$ selection. The $M_r$ values of SDSS galaxies exhibit a strong correlation with subhalo mass indicators (i.e. mass properties and circular velocity properties), in the way that more massive subhaloes host brighter galaxies. Subhalo assembly properties such as $z_{\rm 0.1/0.3/0.5/0.7/0.9}$ and $z_{\rm first/last}$ also demonstrate a correlation with $M_r$. In comparison to $M_r$, correlations between colour and subhalo properties are generally weaker. Moderate correlations ($\sim$ 0.4) are found between colour and mass indicators, and the correlations with subhalo assembly properties are close to zero. Environmental properties such as $\delta_{2.1}$ and $t_{\rm web}$ correlate weakly with both $M_r$ and colour.

The second panel displays the correlations of the original SAM sample using the $z_{\rm lim}$-selected subhaloes. Compared to the findings in SDSS, $M_r$ in the SAM sample correlates weaker with mass indicators, and the correlations with subhalo assembly properties are negligible. In contrast, galaxy colour in the SAM correlates stronger with mass indicators than that in SDSS. This may be the reason that the RF provides a more accurate prediction of galaxy colour in the SAM. Both $M_r$ and colour in the SAM correlate very weakly with halo assembly properties.

In the corresponding shuffled sample shown in the third panel, all the correlation coefficients involving subhalo mass indicators and environmental properties are almost maintained from the original sample due to the shuffling constraints. Since the correlations relating to subhalo assembly properties are weak in the original SAM, the overall correlations between $M_r$ or colour and subhalo properties are essentially unchanged. However, it is important to note that the correlation coefficient captures the correlations between individual pairs of variables instead of the multi-variate dependence. With the shuffling, the multi-variate dependence between colour and subhalo properties experiences small variations, as indicated by the slightly lower $R^2$ after shuffling compared to that before the shuffling.

The fourth panel shows the correlations in the original SAM sample using all subhaloes above $M_{\rm sub}$=10. This sample contains more low-mass subhaloes compared to the $z_{\rm lim}$-selected sample. Compared to the second panel, this sample shows tighter correlations between $M_r$ and mass indicators, as well as the merger tree properties such as $N_{\rm merg}$ and $z_{\rm first/last}$. The correlations between colour and mass indicators are weaker in this sample, while the correlations between colour and subhalo assembly properties are stronger. Positive correlation coefficients suggest that red galaxies tend to reside in early-formed subhaloes. Interestingly, the colour correlates more strongly with late formation stage properties (i.e. $z_{\rm 0.7}$) than those characterising the early formation stage of subhaloes (i.e. $z_{\rm 0.1}$). 

Considering the differences between the second panel and the fourth panel, it is important to acknowledge that generalizing ML models based on the $z_{\rm lim}$-selected subhaloes to the entire ELUCID simulation may introduce biases if the galaxy-subhalo relation also depends on subhalo mass in the real Universe. Observations including more faint galaxies (and thus low-mass subhaloes) such as DESI and constrained $N$-body simulation recovering smaller mass and length scales could be helpful for investigating galaxy-subhalo relations in the low mass range.

\begin{figure*}
    \centering
    \begin{subfigure}[h]{1.1\textwidth}
    \includegraphics[width=\textwidth]{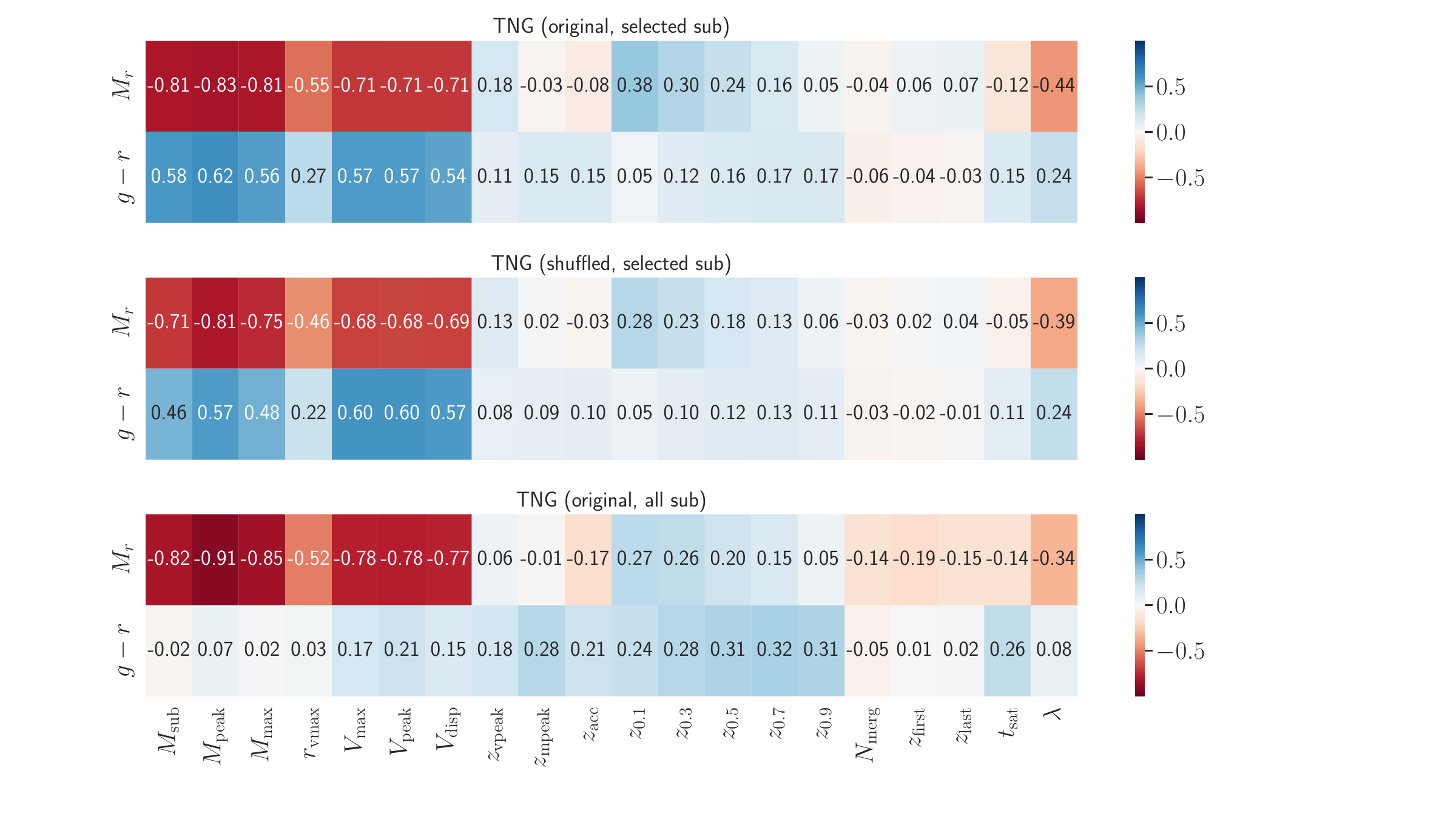}
    \end{subfigure}
\caption{Similar to Figure~\ref{fig:corrcoef-sam}, but for TNG samples. From top to bottom, the samples are original TNG (selected subhaloes), shuffled TNG (selected subhaloes), and original TNG (all subhaloes above $M_{\rm sub}$=10).}
\label{fig:corrcoef-tng}
\end{figure*}

We also conduct the same analysis with TNG galaxies. The top panel of Figure~\ref{fig:corrcoef-tng} presents the correlation coefficients of the selected subhaloes of TNG. $M_r$ is highly correlated with mass indicators and weakly correlated with assembly properties, which are both stronger than those in the SAM. Notably, the $M_r$ correlations with $z_{\rm 0.1} \sim z_{\rm 0.9}$ gradually decrease, suggesting that $M_r$ depends more on the early formation stage than the late formation stage of the subhalo. This trend is absent in the $z_{\rm lim}$-selected SAM sample but is also present in the SDSS sample. Similar to the SAM, the TNG colour moderately correlates with mass indicators and is nearly independent of assembly properties. The second panel shows the results of the shuffled sample. We find again that the shuffling barely affects the correlations related to mass indicators. Additionally, the correlations between $M_r$ and $z_{\rm 0.1} \sim z_{\rm 0.9}$ remain partially intact, along with the gradually decreasing trend. This is possibly due to the shuffling constraint which limits the shuffling within $5\hmpc$ cells, and the subhaloes assembly properties of similar mass may exhibit minimal variations within these cells. 

In the third panel of Figure~\ref{fig:corrcoef-tng} which includes all subhaloes above $M_{\rm sub}$=10, the $M_r$ correlations with mass indicator properties are slightly stronger, and correlations with assembly properties can be both higher (e.g., $N_{\rm merg}$, $z_{\rm first/last}$) and lower (e.g., $z_{\rm 0.1} \sim z_{\rm 0.9}$) compared to the selected sample. With a large amount of low-mass subhaloes, the colour correlations with mass indicators are much lower than those in the selected sample. However, the colour correlations with assembly properties are higher. Overall, the colour-subhalo correlations are weaker in the TNG compared to those in the SAM in all subhalo above $M_{\rm sub}$=10.  

Comparing the results of the SDSS sample in the top panel of Figure~\ref{fig:corrcoef-sam} with the corresponding SAM (second and third panels of Figure~\ref{fig:corrcoef-sam}) and TNG results (top two panels of Figure~\ref{fig:corrcoef-tng}), we find that the $M_r$-subhalo relation in the SDSS is more similar to that in TNG, in terms of the dependence on mass indicators and some of the assembly properties. The SDSS colour-subhalo correlation is weaker than both the SAM and TNG, even after shuffling. So it is possible that the true underlying colour-subhalo connection in SDSS without the mismatch effect is lower than those in the SAM and TNG before shuffling, and baryonic processes such as AGN feedback and other stochastic processes may have significant impacts on SDSS galaxies. Further comparison between the SDSS and TNG galaxies can be carried out with the upcoming ELUCID hydrodynamic simulation (HELUCID, Cui in prep), which can provide new insights into galaxy-subhalo relation in the real Universe.

\section{Summary} 
\label{sec:summary}

Using a catalogue matching SDSS galaxies with ELUCID subhaloes, we employ random forest to predict galaxy magnitude and colour based on a few subhalo properties that characterise subhalo mass, assembly history, and environment. Before training the RF, we select a sample of galaxy-subhalo pairs from the SDSS-ELUCID matched catalogue according to the redshift limitation that corresponds to subhalo mass completeness. This eliminates most of galaxies with subhaloes of ${\rm log}M_{\rm sub}<11$ and a fraction of galaxies with subhaloes of $11<{\rm log}M_{\rm sub}<12$. Training on this selected sample, the RF model can predict the $M_r$ reasonably accurately with an $R^2$ score of $\sim$0.8, with deviations mainly arising from extremely bright and faint galaxies. The prediction can recover the luminosity function and galaxy-matter cross-correlation in the range of $-22<M_r<-18$. Extending the predictions to all ELUCID subhaloes results in slightly larger deviations, especially at the faint end. In contrast, the accuracy of colour prediction is significantly lower, with an $R^2$ score of $\sim$ 0.3. The RF model fails to reproduce the position of the red peak in SDSS $z_{\rm lim}$-selected sample, leading to large deviations in predicted colour values from the true colour. We also train RF models to predict physical galaxy properties such as $M_*$ and sSFR. The prediction performance of $M_*$ is similar to that of the $M_r$, and the prediction performance of sSFR is similar to that of the colour.

One possible explanation for the low accuracy of colour prediction is the difference between the matched subhaloes and the underlying true subhaloes of SDSS galaxies, or in other words, the mismatch between SDSS galaxies and subhaloes. To investigate this effect, we utilise galaxies from a SAM model implemented on ELUCID. We shuffle the galaxies around subhaloes in ${\rm log}M_{\rm peak}$ bins of 0.2 dex and in cubic cells of $5\hmpc$. RF models are trained on the $z_{\rm lim}$-selected subhaloes both before and after the shuffling. Before the shuffling, the colour prediction is reasonable with an $R^2$ of $0.79$, and the bimodal distribution of colour is reproduced. The effect of shuffling lowers the $R^2$ score to $0.66$, but still higher than that of the SDSS sample. 

We also perform the same test using galaxies in TNG300. Since the density field of TNG300 is not directly matched to the SDSS, we select random fractions of subhaloes as a function of ${\rm log}M_{\rm sub}$ to ensure that the selected subhalo sample reproduces the subhalo abundance in the $z_{\rm lim}$-selected subhaloes of SDSS. Before shuffling, the $R^2$ of colour prediction is $0.73$, and it decreases to $0.59$ after shuffling. The impact of shuffling in TNG is comparable to that in the SAM, which slightly lowers the colour-subhalo connection. This finding suggests that the colour-subhalo connection in SDSS may be weaker than both the SAM and TNG, even in the absence of the mismatch effect.

In the end, we measure the Pearson correlation coefficients between $M_r$ or colour and the subhalo properties for SDSS, SAM, and TNG samples. In the SDSS and selected TNG, $M_r$ shows a strong correlation with subhalo mass indicators such as mass properties and circular velocity properties and a weak correlation with subhalo assembly properties. However, these correlations appear weaker in the selected SAM sample. The colour in both selected SAM and TNG correlates moderately with mass indicators and exhibits small dependence on assembly properties, and the correlations between SDSS colour and subhalo properties are weaker than both the SAM and TNG. The shuffling shows minimal effects on the correlation coefficients of both the SAM and TNG samples. In terms of the correlation coefficients, the colour in SDSS also demonstrates a lower connection with subhaloes compared to both the SAM and TNG, taking the mismatch effect into consideration.

We also show that the correlation coefficients in SAM and TNG depend on the subhalo mass. Including more low-mass subhaloes, the $M_r$ correlation coefficients increase in both SAM and TNG, especially those with mass indicators. The colour correlations with subhalo assembly properties also increase, but those with mass indicators decrease, especially those in the TNG. It is possible that the galaxy-subhalo correlation in SDSS also depends on subhalo mass. Appropriate care should be taken when generalizing studies of galaxy-halo connection from SDSS-like subhaloes to a broader range of halo masses.

The results above suggest that it is reasonable to learn the $M_r$-subhalo relation with machine learning using a galaxy-subhalo matched catalogue built on constrained simulation, but it is difficult to capture the colour-subhalo relation. The HELUCID simulation which includes baryonic particles in addition to the dark matter particles will be available in the future. It is helpful for investigating the difference between simulated galaxies and real galaxies on a one-to-one level and provides new insights into galaxy-halo relations and galaxy formation and evolution. Advanced surveys such as DESI which include more faint galaxies are also important in resolving the issue of SDSS on the galaxy-halo relations in the low-mass end.

\section*{Acknowledgements}
This work is supported by the National Science Foundation of China (grant Nos. 11833005, 11890692, 11621303), 111 project No. B20019, and Shanghai Natural Science Foundation, grant No.19ZR1466800. We acknowledge the science research grants from the China Manned Space Project with No.CMS-CSST-2021-A02. XX acknowledges the support from Shanghai Post-doctoral Excellence Program (2021231) and China Postdoctoral Science Foundation (2022M712085). YZ acknowledges the support from the National Science Foundation of China (grant No. 12273088)

\section*{Data Availability}
The ELUCID simulation and the SDSS-ELUCID matched catalogue used in this work can be accessed at \url{https://gax.sjtu.edu.cn/data/ELUCID.html}. TNG simulation data can be accessed at \url{http://www.TNG-project.org}.

\bibliographystyle{mnras}
\bibliography{output}

\appendix
\section{Predicting $M_*$ and sSFR}
\label{sec:appendix}

\begin{figure*}
    \centering
    \begin{subfigure}[h]{0.99\textwidth}
    \includegraphics[width=\textwidth]{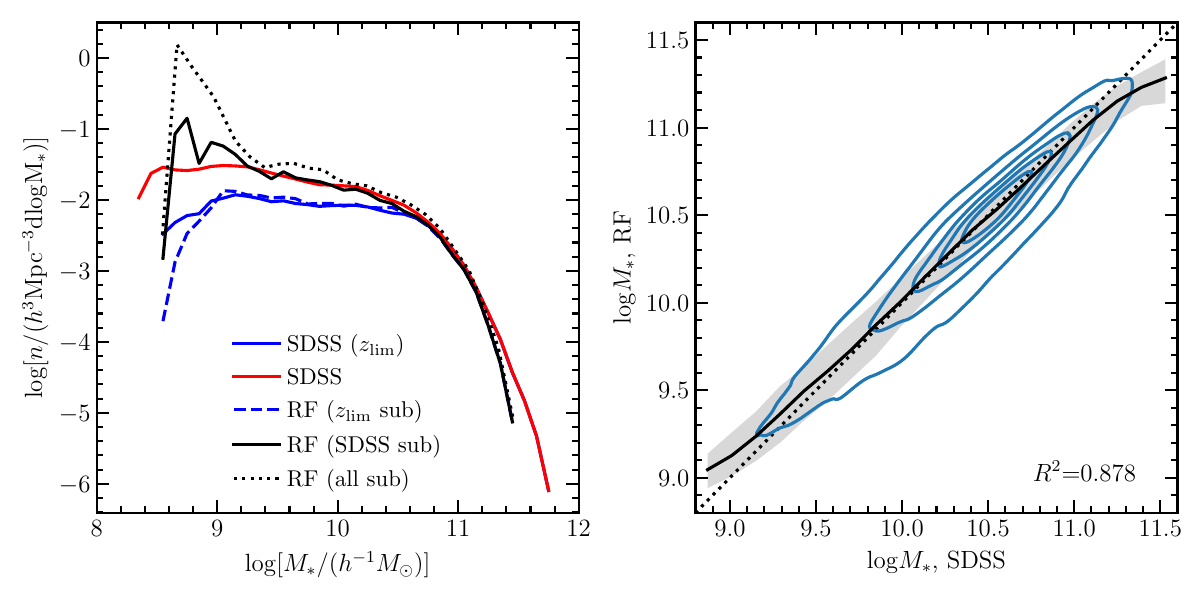}
    \end{subfigure}
\caption{Similar to Figure~\ref{fig:sdss-mr}, but for stellar mass.}
\label{fig:sdss-mstar}
\end{figure*}

\begin{figure*}
    \centering
    \begin{subfigure}[h]{0.99\textwidth}
    \includegraphics[width=\textwidth]{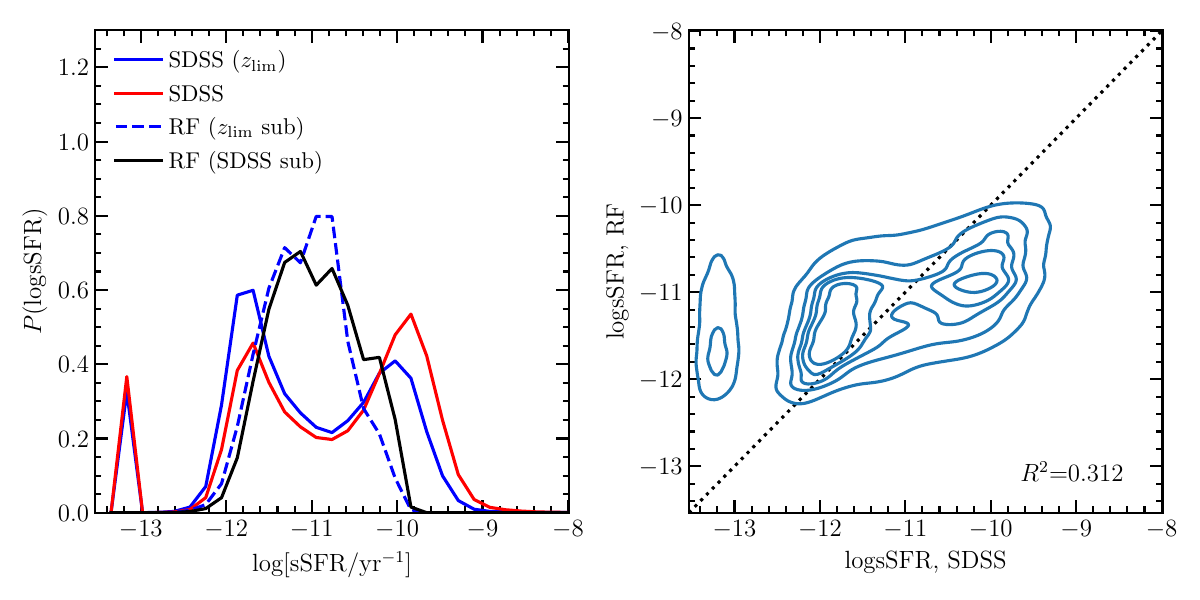}
    \end{subfigure}
\caption{Similar to Figure~\ref{fig:sdss-color}, but for specific star formation rate.}
\label{fig:sdss-ssfr}
\end{figure*}

Similar to those in Section~\ref{sec:mrpred} and Section~\ref{sec:colorpred}, we also train RF models to predict stellar mass $M_*$ and sSFR. The $M_*$ of SDSS galaxies is computed according to \citet{Bell2003} based on the stellar mass-to-light ratio and galaxy colour (see \citet{Yang2018} for details). The RF results of $M_*$ are present in Figure~\ref{fig:sdss-mstar}. The left panel shows the stellar mass function (SMF), where the red (blue) solid indicates the SDSS measurement of all SDSS-ELUCID matched ($z_{\rm lim}$-selected) galaxies. To measure the SMF, we compute the redshift completeness limit based on stellar mass according to Equation (8) of \citet{Yang2018} and select galaxies with redshift above the limit. Similar to that in Figure~\ref{fig:sdss-color}, SMF of the $z_{\rm lim}$-selected sample is lower than that of the SDSS-ELUCID sample below ${\rm log} M_*<10.6$. RF prediction of the $z_{\rm lim}$ (all SDSS-ELUCID) subhaloes is represented by the blue dashed (black solid). We find reasonable agreement between prediction and observation, except for the high-mass and the low-mass ends. Applying the RF on all ELUCID subhaloes (black dashed), a bump presents at the low-mass end, similar to that in Figure~\ref{fig:sdss-mr}. 

The right panel directly compares the SDSS $M_*$ and the predicted $M_*$ for the $z_{\rm lim}$ sample. The $R^2$ of the prediction is 0.878, which is slightly higher than that of the $M_r$ prediction. This may be attributed to the neighbourhood abundance matching between the SDSS galaxies and ELUCID subhaloes, which links the stellar mass to $M_{\rm peak}$. 

Figure~\ref{fig:sdss-ssfr} displays the RF results for sSFR, which is the ratio of SFR to $M_*$. The SFR in our sample is obtained from the estimation of \citet{Brinchmann2004}. In the left panel, the blue and red solid indicate the distribution of sSFR of the $z_{\rm lim}$-selected and the full SDSS-ELUCID sample, respectively. The distributions exhibit peaks around $\rm log sSFR=-11.8$ and $\rm log sSFR=-10$, as well as a population of zero star formation which is assigned with sSFR of $\rm log sSFR=-13.2$. However, these peaks are not reproduced by the predictions. The right panel also shows that the prediction deviates significantly from the true values with an $R^2$ of 0.3, similar to the colour prediction. This illustrates again that the correlations between star-forming activity and subhalo properties are weak in SDSS.

\bsp	
\label{lastpage}
\end{document}